# Ultrafast electron-electron scattering in metallic phase of 2H-NbSe$_2$ probed by high harmonic generation

K. S. Takeda[1], K. Uchida[1,*], K. Nagai[1], S. Kusaba[1], and K. Tanaka[1,*]

[1] *Department of Physics, Graduate School of Science, Kyoto University, Kyoto, Kyoto 606-8502, Japan*

*e-mail: uchida.kento.4z@kyoto-u.ac.jp, kochan@scphys.kyoto-u.ac.jp

## Abstract

Electron-electron scattering on the order of a few to tens of femtoseconds plays a crucial role in the ultrafast electron dynamics of conventional metals. When mid-infrared light is used for driving and the period of light field is comparable to the scattering time in metals, unique light-driven states and nonlinear optical responses associated with the scattering process are expected to occur. Here, we use high-harmonics spectroscopy to investigate the effect of electron-electron scattering on the electron dynamics in thin film 2H-NbSe$_2$ driven by a mid-infrared field. We observed odd-order high harmonics up to 9th order as well as a broadband emission from hot electrons in the energy range from 1.5 to 4.0 eV. The electron-electron scattering time in NbSe$_2$ was estimated from the broadband emission to be almost the same as the period of the mid-infrared light field. A comparison between experimental results and a numerical calculation reveals that a kind of "resonance" between scattering and driving enhances the non-perturbative behavior of high harmonics in metals, causing a highly non-equilibrium electronic state corresponding to several thousand Kelvin.

Electron dynamics driven by an intense light field induce novel nonlinear optical phenomena and sometimes result in non-trivial light-induced phase transitions in solids [1-3]. An understanding of the dynamics is crucial to not only fundamental optical science but also the development of applications such as ultrafast manipulation of material properties and femtosecond-laser processing [4-7]. In dilute atomic or molecular systems, where strong-field physics have been intensively studied, electron motion is fully ballistic and coherent without any scattering. On the other hand, in solids, where interactions involve multiple degrees of freedom, scattering by other electrons or other degrees of freedom occurs under driving. When the typical scattering time scale is longer than the period of the driving field, the electron motion can be regarded as coherent and ballistic, as in atoms [2,8-11]. On the other hand, when the typical scattering time is shorter than the driving period, scattering disrupts the ballistic electron motion (called a drift-diffusive regime) [12] and prevents the formation of non-equilibrium states. In the intermediate regime where the scattering time is comparable to the driving-field period [13], the interplay between driving and scattering causes electrons to absorb a huge amount of energy from the driving field and the electron distribution moves far from the equilibrium one. Nonlinear optical response reflecting a scattering process unique to solids is also expected to occur in this regime. So far, however, experimental studies on ultrafast electron dynamics in solids under driving fields often have neglected the effects of scattering and dissipation processes.

Metals under a strong mid-infrared (MIR) driving field provide a fascinating playground for studying light-driven states and nonlinear optical processes associated with scattering. The electron-electron (e-e) scattering time in metals is typically about 1-10 fs [14], and this value is comparable to the period of MIR light, i.e., the intermediate regime discussed above. Moreover, in metals, we can assume the simplest condition, i.e., intraband motion of electrons within a single band. This contrasts with the circumstances of semiconductors and semimetals, where electrons in principle move across multiple bands and their dynamics are quite complicated. However, so far, extremely nonlinear optical processes such as high order harmonic generation (HHG) have mainly been studied in semiconductors and insulators [15,16], whereas only a few studies have targeted metals [17]. This is because the electric field is strongly screened inside metals, making it difficult to apply a strong MIR electric field to electrons in bulk metals. It is still a challenging problem to observe the electron dynamics in the intermediate regime.

In this study, we performed nonlinear emission measurements in a thin film of metallic-phase of 2H-NbSe$_2$ driven by intense MIR light in order to clarify the electron dynamics with ultrafast dissipation under driving. We observed high harmonics from 5$^{th}$ to 9$^{th}$ order as well as a broadband emission from 1.5 eV to 4.0 eV that has not been observed before in semiconductors and insulators. We attribute the nonlinear broadband emission to incoherent thermal radiation from hot electrons reaching 4000 K. These results indicate that the electrons in the metal are driven in reciprocal space with a large amplitude and strongly scattered at the same time, forming a high-temperature electron distribution immediately after driving. Our numerical results reveal that the e-e scattering time is comparable to the period of the driving field and that cooperation between driving and scattering leads to unique extreme nonlinear optical properties in metals.

The target sample was a thin film of metallic 2H-NbSe$_2$. Since 2H-NbSe$_2$ is a van der Waals layered material, we could obtain a sample with a thickness of 20 nm via the mechanical exfoliation method [18], which is sufficiently thinner than the penetration length of the excitation light (about 80 nm) [19]. Inside such a thin film, screening of the electric field, which occurs in bulk metals and often suppresses HHG signals [17], can be neglected. In addition, 2H-NbSe$_2$ has a simple quasi-two-dimensional band structure near the Fermi level, which can be easily put into a model [20,21].

Figure 1(a) shows a schematic diagram of the experimental setup. MIR pulses (center wavelength (photon energy): 4.8 μm (0.26 eV), pulse duration: 60 fs) were generated by using differential frequency generation between the signal and idler outputs from an optical parametric amplifier (OPA, TOPAS-C, Light Conversion). The input of the OPA was part of the output from a Ti: sapphire regenerative amplifier (800 nm center wavelength, 35 fs pulse duration, 1 kHz repetition rate, and 1 mJ pulse energy). The polarization of the MIR light was controlled with a wire-grid polarizer and liquid crystal retarder (LCC1113-MIR, Thorlabs), and it was focused on the sample. Since the sample was a thin film, the experiments were conducted in a transmission configuration. The spot size of the MIR pulses was about 60 μm (full width at half maxima (FWHM)). To suppress damaging the samples during the irradiation, they were mounted in a vacuum vessel (<10$^{-3}$ Pa). The maximum intensity of the MIR light was set to be 77 GW/cm$^2$ in a vacuum because a sample was damaged above this intensity. The corresponding electric field inside the sample $E_{MIR}$ was estimated to be 3.8 MV/cm. A pair of CaF$_2$ windows was used to compensate for the group delay dispersion that accumulated by the light passing through the optics, such as the liquid crystal retarder. We resolved the polarization of the nonlinear emissions from the sample by using a wire-grid polarizer and detected their spectra by using a spectrometer equipped with a cooled-CCD camera.

The upper panel of Fig. 1(b) shows typical emission spectra from 2H-NbSe$_2$ at an MIR intensity of 77 GW/cm$^2$ ($E_{MIR}$ = 3.8 MV/cm). One can see the peaks of the 5$^{th}$-, 7$^{th}$-, and 9$^{th}$-order harmonics in the emission that is polarized parallel to the MIR field (the blue solid line). There is no even-order harmonics, reflecting the centrosymmetric crystal structure of 2H-NbSe$_2$ [22]. Since the number of carriers generated through interband Zener tunneling in our experiment was estimated to be an order of magnitude smaller than that of the original free carriers [23,24], the high harmonics in 2H-NbSe$_2$ should have mainly come from the intraband motion of the original free carriers [17]. The observation of harmonics up to 9$^{th}$ order indicates that the electron distribution is largely displaced from the equilibrium position in the Brillouin zone under the driving condition. Besides the high harmonics signals, a broadband emission was also observed in the photon energy range from 1.5 eV to 4.0 eV in both the parallel and perpendicular components.

Figure 1(c) shows the polarization state of the 5$^{th}$ (orange circles) and 7$^{th}$ (blue squares) harmonics and the broadband emission (green triangles). The horizontal axis corresponds to the polarization direction of the excitation MIR light. The harmonic signals are well fitted with a cosine-squared function with a peak intensity along the horizontal axis, indicating the polarization of the harmonic signals is parallel to that of the incident MIR light. The broadband emission, on the other hand, is unpolarized. Hence, we can extract the high harmonic signal by subtracting the perpendicular component from the parallel component as shown in the bottom panel of Fig. 1(b).

Such a broadband emission has not been observed in HHG measurements on semiconductors and has only been reported for semimetallic graphene and metallic carbon nanotubes [25,26]. Several studies reported that photoexcitation of metals and semimetals causes broadband emissions similar to our observation [27–30]. In these studies, the broadband emission was attributed to radiation originating from hot electrons that formed after electrons with high excess energy were created through near-infrared or visible light excitation of interband transitions. The broadband emission observed in our study should also have originated from hot electrons. The difference between our experiment and the previous studies is that the hot electrons formed through intense MIR driving of intraband electron motion, where the MIR photon energy is far below the bandgap energy of 2H-NbSe$_2$ [19].

Assuming that the electron distribution after driving can be described by an effective temperature, the high energy side of the broadband emission spectrum can be fitted to the Planck distribution (Fig, 1(d)) [27]. The effective temperature estimated from this fit is 3800 K, and it is considered to be a transient value just after driving. This is because the effective electron temperature decreases over time through the electron-phonon coupling, and an electron distribution with a low effective temperature does not contribute to the broadband emission in the visible range [19]. The fact that the estimated temperature is far above the melting temperature of bulk 2H-NbSe$_2$ in vacuum also supports the conclusion that the broadband emission originates from a hot-electron sub-system before thermalizing with the lattice. The coexistence of high-order harmonics and a thermal emission in 2H-NbSe$_2$ suggests unique electron dynamics are at play in metals in the intermediate regime.

As mentioned above, the scattering process of the driven electrons should play a crucial role in the high-temperature electron distribution. Among the processes in metals, e-e scattering, which conserves the total energy of electrons, is the dominant contribution within the cycle of the driving field (16 fs). Figure 2(a) shows schematic diagrams of the energy transfer from a laser electric field to electrons in metals through intraband e-e scattering. When electrons follow the acceleration theorem without scattering, although their distribution is highly displaced during the driving, the final distribution after driving should be the same as the initial one, i.e., electrons absorb no energy from the driving field (left-hand side of Fig. 2(a)). A previous study on HHG in metallic TiN film was performed in this regime [17]. In that study, visible light, whose period is shorter than the scattering time, was used for the excitation, and the most of the results could be accounted for by electron dynamics according to the acceleration theorem. On the other hand, when scattering is involved in the driven electron dynamics, the displaced electrons are scattered in reciprocal space, resulting in a high-temperature electron distribution after the driving (right-hand side of Fig. 2(a)).

To verify the above scenario, let us consider the Boltzmann transport equation [31]:

$$\frac{\partial f}{\partial t} - \frac{e}{\hbar}\mathcal{E}(t)\cdot\nabla_{\boldsymbol{k}}f = \Sigma_{\text{scatter}}, \qquad (1)$$

where $f = f(\boldsymbol{k},t)$ is the electron distribution function, $\boldsymbol{k}$ is the crystal momentum, $e(>0)$ is the electron charge, $\hbar$ is the Planck constant, and $\mathcal{E}(t)$ is the MIR electric field. For simplicity, we ignore the spatial diffusion term in the Boltzmann equation. In a simulation, we took $\boldsymbol{A}(t) = -\int_{-\infty}^{t}dt'\mathcal{E}(t')$ as the Gaussian pulse and used the experimental values for the pulse width (FWHM) and center frequency. The initial conditions were $f(\boldsymbol{k},-\infty) = f_{\text{F}}(\boldsymbol{k},T=300\ K)$, where $f_{\text{F}}(\boldsymbol{k},T)$ is the Fermi distribution function at temperature $T$. From $f(\boldsymbol{k},t)$ one can calculate the electronic current as

$$\boldsymbol{j}(t) \propto -e\int_{\text{BZ}}d\boldsymbol{k}\boldsymbol{v}(\boldsymbol{k})f(\boldsymbol{k},t) = -\int_{\text{BZ}}d\boldsymbol{k}\frac{e}{\hbar}\nabla_{\boldsymbol{k}}E(\boldsymbol{k})f(\boldsymbol{k},t), \qquad (2)$$

where $E(\boldsymbol{k})$ is the band structure of 2H-NbSe$_2$. For simplicity, we used a band structure calculated from the two-dimensional tight binding model with three d-orbitals [19-21]. From the calculated bands, we took the band that crossed the Fermi energy as $E(\mathbf{k})$. The HHG spectrum is obtained from the Fourier transform of the current and given by $I_{HH}(\omega) \propto \omega|\tilde{\mathbf{j}}(\omega)|^2$.

To take e-e scattering into account, let us consider the simplest scattering process, i.e., back-scattering, which causes a large modification to the electron distribution. We used the following phenomenological form, which is extended from the one used in Ref. 32 [32,19].

$$\Sigma_{\text{scatter}} = \sum_g \gamma_g \{f(\mathbf{k},t) - f(g\mathbf{k},t)\}, \qquad (3)$$

where g is the symmetric operation determined by the crystal symmetry of 2H-NbSe$_2$ (point group of D$_{6h}$) and $\gamma_g$ is the scattering rate of each scattering path labeled by the symmetry operation g [19]. This form includes the Umklapp process and effectively describes the back-scattering of electrons. A real system with a complicated band structure would have other scattering paths for electrons, which cannot be described by back scattering. Here, for simplicity, we will assume that only the back-scattering channels are relevant in the time scale of interest (~10 fs) and that the scattering rate is the same for all paths in Eq. (3) ($\gamma_g = \gamma$). We used the Fourier expansion method for the numerical calculation, and it enabled the electronic current and electron distribution to be calculated with high accuracy at a low computational cost [19].

Figure 2(c) shows the estimated total energy of electrons after driving as a function of the scattering time for $E_{MIR} = 3.8$ MV/cm [19]. Importantly, the total energy does not monotonically decrease as the scattering time increases and has a maximum value at around 10 fs, which is comparable to the period of the driving field (16 fs). This indicates a kind of "resonance" between the scattering time and the period of the driving field, which can be explained as follows. When the scattering is much longer than the cycle of the driving field, electrons follow the acceleration theorem and do not absorb energy much, as explained previously. On the other hand, when the scattering time is much shorter than the period, electrons are scattered immediately before they move a long way in reciprocal space; thus, the changes in the electron distribution are negligible. In other words, a highly non-equilibrium electronic distribution can be realized only in the intermediate regime where the scattering time is comparable to the cycle of the driving field. The estimated temperature of 3800 K for $E_{MIR} = 3.8$ MV/cm indicates that our experimental condition is indeed in the intermediate regime.

We estimated the typical scattering time in 2H-NbSe$_2$ by using the electron temperature value of 3800 K obtained from the broadband emission [19]. As shown in Fig. 2(c), there are two possible scattering times corresponding to an electron temperature of 3800 K. We chose the longer one since high harmonic emission was strongly suppressed for the shorter time in our simulation. By taking all scattering paths (11 paths) into account, the scattering time is estimated to be $\gamma^{-1}/11 = 14$ fs ($\gamma^{-1} = 155$ fs). The estimated value is consistent with typical e-e scattering times in metals [14] and comparable to the period of the MIR electric field (16 fs). This indicates that our experimental conditions satisfy the intermediate regime discussed above.

To check whether the observed HHG results can be reproduced by our model using the scattering time estimated above ($\gamma^{-1} = 155$ fs), we compared the experimental values with numerical HHG results. Figure 3(a) shows the experimental (red solid line) and calculated (gray dashed line) high harmonics spectra with $E_{MIR} = 3.8$ MV/cm. The numerical results reproduce the observed HHG spectrum, indicating that the scattering time estimated from the incoherent broadband emission is consistent with the scattering time in the coherent electron dynamics reflected in the observed high harmonic emission.

Figure 3(b) shows the 5$^{th}$- and 7$^{th}$-order harmonics intensity as a function of the polarization direction of MIR light with $E_{MIR} = 3.8$ MV/cm. Here, 2H-NbSe$_2$ shows an almost isotropic crystal orientation dependence. This is in contrast to the rather anisotropic crystal orientation dependence of the harmonics in several crystalline solids reflecting inter-atomic bonding and the band structure [33–40]. This feature is also reproduced by the numerical calculation with $E_{MIR} = 3.8$ MV/cm and $\gamma^{-1} = 155$ fs. In metals, all electrons below the Fermi level contribute to the nonlinear current (high harmonics). In 2H-NbSe$_2$, integrating the contributions cancels out the anisotropic response of each electron, resulting in an isotropic response as in the experiments. The numerical calculation predicts that the anisotropy of high harmonics strongly depend on the Fermi level, which is a characteristic feature of HHG in metals [19].

To clarify the non-perturbative nature of HHG in 2H-NbSe$_2$, we plot the ratio of the *n*-th order harmonics intensity to the *n*-th power of MIR intensity ($I_n/I_{\text{MIR}}^n$) as a function of MIR intensity in Fig. 3(c). This ratio should be constant for sufficiently weak MIR intensity because high harmonics signals obey a power law ($I_n \propto I_{\text{MIR}}^n$) in the perturbative regime. When the light-matter interaction enters the non-perturbative regime as the MIR intensity increases, the ratio starts to depend on the MIR intensity, reflecting the electron dynamics unique to each material [15]. In our experiments, both the 5[th] and 7[th] harmonics show strong deviations from constant values, indicating a non-perturbative nonlinear optical process above 2 MV/cm. The numerical results for e-e scattering ($\gamma^{-1} = 155$ fs) reproduce the saturation seen in the experimental results, while the results without scattering are only weakly saturated. Overall, these results indicate that the scattering time estimated from the broadband emission is valid and that e-e scattering plays a crucial role in the non-perturbative nature of HHG in 2H-NbSe$_2$.

To investigate the effect of e-e scattering on the HHG process in more detail, we calculated the MIR intensity dependence of the high harmonics by changing the scattering time as shown in Fig. 3(d). The orange shaded area is the range of variation in the experiment. The deviation from perturbation theory monotonically increases with increasing MIR intensity and decreasing scattering time. This behavior can be understood as follows. When the MIR electric field is sufficiently weak, the change in the electron distribution from equilibrium is small, limiting the number of available scattering channels. As the MIR electric field increases, the electron distribution becomes more displaced from equilibrium, leading to an increase in available scattering channels. This increase in the effective scattering rate with the MIR electric field suppresses the high harmonic signal originating from coherent electron motion. Since such a field-dependent scattering process should be universal in solids [13], the above scenario may substantially contribute to the non-perturbative nonlinearity of the optical response in a wide variety of systems, including metals and semiconductors.

In conclusion, we observed coherent high harmonic signals and an incoherent broadband emission induced by intense MIR driving in the metallic phase of 2H-NbSe$_2$. Since the incoherent broadband emission associated with HHG has not been observed in semiconductors and insulators, our observation indicates that the electron dynamics under strong MIR electric fields are unique to metals. We conclude that the broadband emission originates from hot electrons, whose effective temperature reaches around 4000 K. The numerical calculation including e-e scattering reproduced both the high harmonic spectrum and high electron temperature simply by tuning the e-e scattering time. The calculation revealed that there is a kind "resonant" condition between the scattering time and driving field period for electrons to energy absorb in the intermediate regime. This "resonant" condition is the cause of the unique nonlinear optical properties, such as scattering-induced non-perturbative nonlinearity, found in the HHG measurements. Our research proves that metals under a strong MIR field are suitable for studying the effect of ultrafast scattering on driven-electron systems in the intermediate regime. Our work provides a new method to access non-equilibrium dynamics with dissipation.

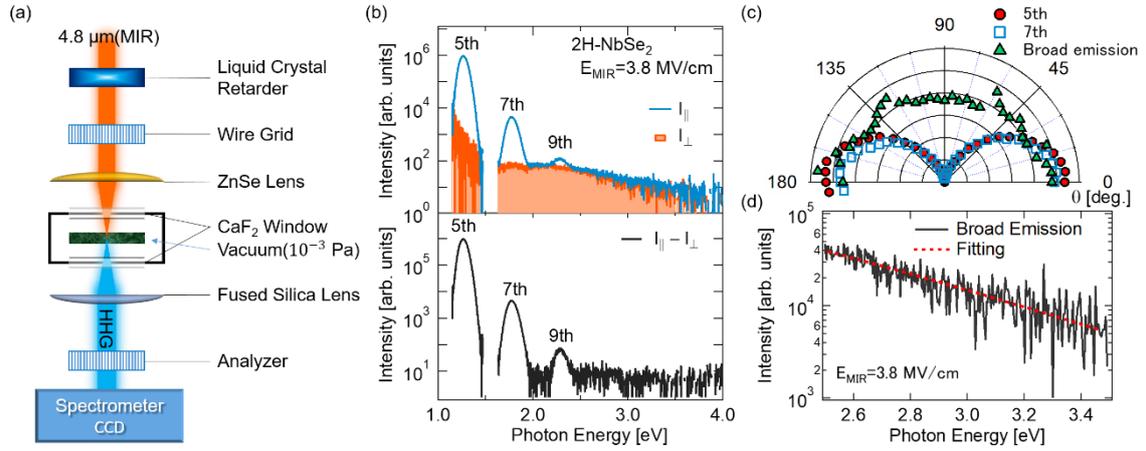

FIG. 1. (a) Experimental setup of HHG measurement. (b) Typical HHG spectra in 2H-NbSe$_2$. Here, the MIR electric field inside the sample $E_{MIR}$ is estimated to be 3.8 MV/cm (corresponding to an MIR intensity of 77 GW/cm$^2$ in vacuum). The blue line in the upper panel shows the emission polarized parallel to the MIR electric field. The orange shaded area shows the perpendicular component. The lower panel is the high harmonics spectrum obtained by subtracting the perpendicular component from the parallel component. (c) High harmonic intensity as a function of the transmission axis angle θ of the analyzer. Here, θ = 0° corresponds to the MIR electric field direction. Red circles and blue rectangles show the 5$^{th}$ and 7$^{th}$ harmonics, respectively. Green triangles show the broadband emission integrated over the region between 2.5 and 4 eV. (d) Broadband emission spectrum above 2.5 eV with $E_{MIR}$ = 3.8 MV/cm. The red dotted line represents the fitting using the Planck distribution. The effective temperature estimated from the fitting is about 3800 K.

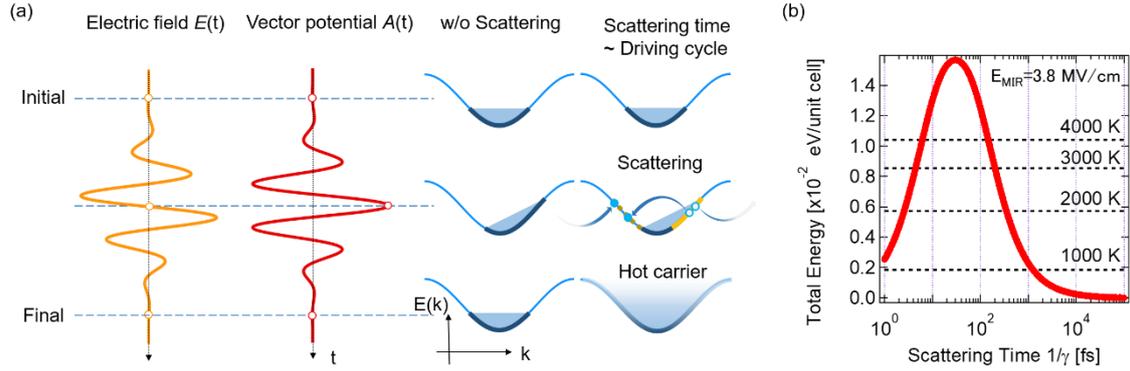

FIG. 2. (a) Schematic time evolution of the carrier distribution. Left panel shows the temporal profile of MIR electric field and the corresponding vector potential. Middle and right panels show the electron distribution in a one-dimensional cosine band at each time without and with the scattering effect, respectively. According to the acceleration theorem, the canonical momentum of electron is temporally changed in proportion to the vector potential (middle panel). In this situation, the electron distribution returns to the original distribution before driving. The scattering process causes the electron distribution to be different from the original one (right panel). (c) Total energy of electron distribution of 2H-NbSe$_2$ after MIR-pulse excitation with $E_{MIR} = 3.8$ MV/cm. The red line shows the total energy of the electron distribution that is asymptotic to the total energy before the excitation in the limit $\gamma^{-1} \to \infty$. Black dotted lines show the total energies of electrons for several temperatures.

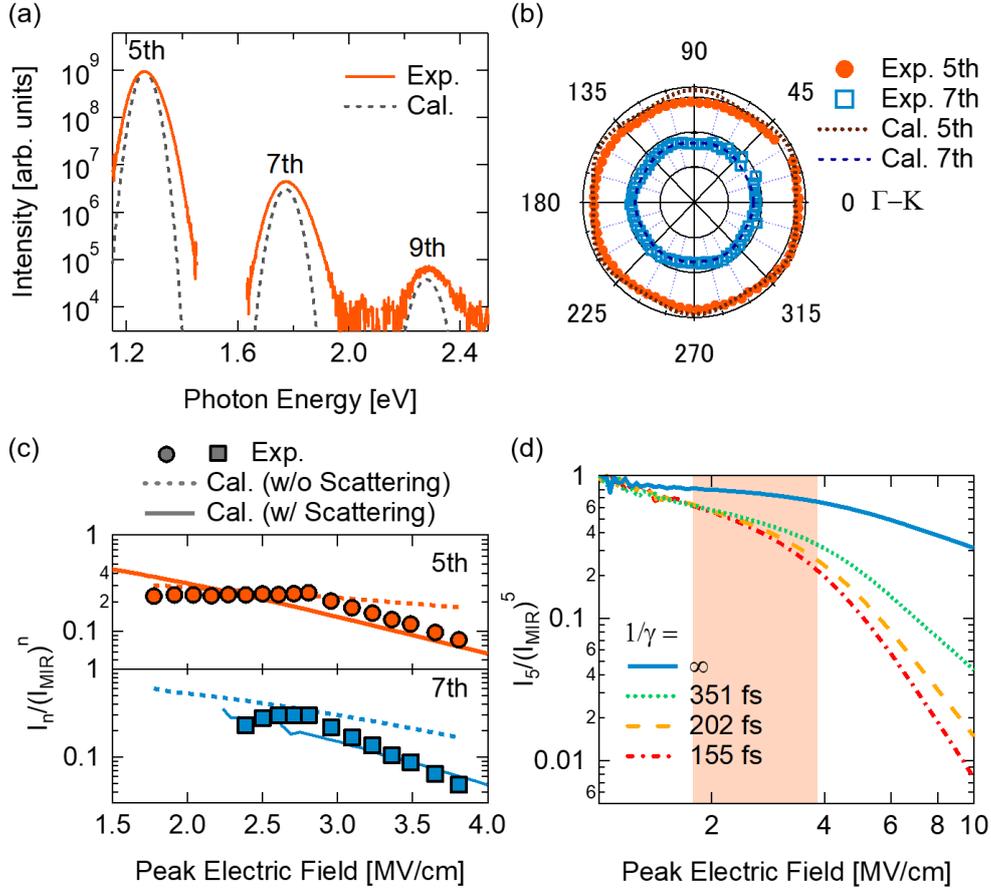

FIG. 3. Comparison of experimental and calculated results. (a) High harmonic spectra with $E_{MIR} = 3.8$ MV/cm. Orange solid and black dashed lines show the experimental and numerical results, respectively. The simulation used a scattering time of $\gamma^{-1} = 155$ fs. (b) Crystal orientation dependence of the 5th and 7th harmonics. Orange solid circles and blue open squares show the experimental results for the 5th and 7th harmonics, respectively. The brown dotted and dark-blue solid lines show the numerical results ($E_{MIR} = 3.8$ MV/cm and $\gamma^{-1} = 155$ fs) of the 5th and 7th harmonics, respectively. 0 degrees corresponds to the Γ-K (zigzag) direction. For clarity, the intensity of the 7th harmonics are magnified. (c) Ratio of *n*-th order harmonic intensity to the *n*-th power of MIR intensity ($I_n/I_{MIR}^n$) as a function of MIR electric field. Symbols (circles and squares), solid, and dotted lines

respectively show the experimental results, numerical results with and without scattering effect ($\gamma^{-1} = $ 155 fs and $ = \infty$). The upper panel shows the 5$^{th}$ harmonic and lower panel shows the 7$^{th}$. Here, all the data are normalized by the value at $E_{MIR} = $ 2.5 MV/cm for clarity. (d) Ratio of the 5$^{th}$ harmonic intensity to the fifth power of MIR intensity ($I_5/I_{MIR}^5$) as a function of MIR electric field at certain scattering times. The blue solid line shows the results without the scattering effect. Green dotted, yellow dashed, and red dashed-dotted lines show the results with a scattering time of 351 fs, 202 fs, and 155 fs, respectively. Here, all the data are normalized by the value at $E_{MIR} = $ 2.5 MV/cm. The shaded area shows the range of the MIR electric field in (c).

[1] T. Oka, S. Kitamura, *Floquet Engineering of Quantum Materials*. Annu. Rev. Condens. Matter Phys. **10**, 387 (2019).

[2] S. Y. Kruchinin, F. Krausz, and V. S. Yakovlev, *Colloquium: Strong-Field Phenomena in Periodic Systems*. Rev. Mod. Phys. **90**, 021002 (2018).

[3] A. de la Torre, D. M. Kennes, M. Claassen, S. Gerber, J. W. McIver, and M. A. Sentef, *Colloquium: Nonthermal pathways to ultrafast control in quantum materials.* Rev. Mod. Phys. **93**, 41002 (2021).

[4] M. Malinauskas, A. Žukauskas, S. Hasegawa, Y. Hayasaki, V. Mizeikis, R. Buividas, and S. Juodkazis, *Ultrafast laser processing of materials: from science to industry*. Light Sci. Appl. **5**, e16133 (2016).

[5] F. Langer, C. P. Schmid, S. Schlauderer, M. Gmitra, J. Fabian, P. Nagler, C. Schüller, T. Korn, P. G. Hawkins, J. T. Steiner, U. Huttner, S. W. Koch, M. Kira, and R. Huber, *Lightwave valleytronics in a monolayer of tungsten diselenide*. Nature **557**, 76 (2018).

[6] Á. Jiménez-Galán, R. E. F. Silva, O. Smirnova, and M. Ivanov, *Lightwave control of topological properties in 2D materials for sub-cycle and non-resonant valley manipulation*. Nat. Photonics **14**, 728 (2020).

[7] J.-Y. Shan, M. Ye, H. Chu, S. Lee, J.-G. Park, L. Balents, and D. Hsieh, *Giant modulation of optical nonlinearity by Floquet engineering*. Nature **600**, 235 (2021).

[8] W. Kuehn, P. Gaal, K. Reimann, M. Woerner, T. Elsaesser, and R. Hey, *Coherent Ballistic Motion of Electrons in a Periodic Potential*. Phys. Rev. Lett. **104**, 146602 (2010).

[9] F. Langer, M. Hohenleutner, C. P. Schmid, C. Poellmann, P. Nagler, T. Korn, C. Schüller, M. S. Sherwin, U. Huttner, J. T. Steiner, S. W. Koch, M. Kira, and R. Huber, *Lightwave-driven quasiparticle collisions on a subcycle timescale*. Nature **533**, 225 (2016).

[10] T. Higuchi, C. Heide, K. Ullmann, H. B. Weber, and P. Hommelhoff, *Light-field-driven currents in graphene*. Nature **550**, 224 (2017).

[11] J. W. McIver, B. Schulte, F.-U. Stein, T. Matsuyama, G. Jotzu, G. Meier, and A. Cavalleri, *Light-induced anomalous Hall effect in graphene*. Nat. Phys. **16**, 38 (2020).

[12] A. Jüngel, *Transport Equations for Semiconductors*. 1st ed. (Springer, Berlin, Heidelberg, 2009).

[13] J. B. Gunn, *Instabilities of current in III–V semiconductors*. IBM J. Res. Dev. **8**, 141 (1964).

[14] N. W. Ashcroft and N. D. Mermin, *Solid State Physics*. (Saunders College Publishing, 1976).

[15] S. Ghimire and D. A. Reis, *High-harmonic generation from solids*. Nat. Phys. **15**, 10 (2019).

[16] J. Park, A. Subramani, S. Kim, and M. F. Ciappina, *Recent trends in high-order harmonic generation in solids*. Adv. Phys. :X **7**, 20032 (2022).

[17] A. Korobenko, S. Saha, A. T. K. Godfrey, M. Gertsvolf, A. Yu. Naumov, D. M. Villeneuve, A. Boltasseva, V. M. Shalaev, and P. B. Corkum, *High-harmonic generation in metallic titanium nitride*. Nat. Commun. **12**, 4981 (2021).

[18] K. Novoselov, A. K. Geim, S. V. Morozov, D. Jiang, Y. Zhang, S. Dubonos, I. Grigorieva, and A. A. Firsov, *Electric Field Effect in Atomically Thin Carbon Films*. Science **306**, 666 (2004).

[19] See Supplemental Information for additional details.

[20] G.-B. Liu, W.-Y. Shan, Y. Yao, W. Yao, and D. Xiao, *Three-band tight-binding model for monolayers of group-VIB transition metal dichalcogenides*. Phys. Rev. B **88**, 085433 (2013).

[21] D. Möckli and M. Khodas, *Robust parity-mixed superconductivity in disordered monolayer transition metal dichalcogenides*. Phys. Rev. B **98**, 144518 (2018).

[22] J. A. Wilson and A. D. Yoffe, *The transition metal dichalcogenides discussion and interpretation of the observed optical, electrical and structural properties*. Adv. Phys. **18**, 193 (1969).

[23] E. O. Kane, *Zener tunneling in semiconductors*. J Phys Chem Solids **12**, 181 (1960).

[24] W. Kuehn, P. Gaal, K. Reimann, M. Woerner, T. Elsaesser, and R. Hey, *Terahertz-induced interband tunneling of electrons in GaAs*. Phys. Rev. B **82**, 075204 (2010).

[25] N. Yoshikawa, T. Tamaya, and K. Tanaka, *High-harmonic generation in graphene enhanced by elliptically polarized light excitation*. Science **356**, 736 (2017).

[26] H. Nishidome, K. Nagai, K. Uchida, Y. Ichinose, Y. Yomogida, Y. Miyata, K. Tanaka, and K. Yanagi, *Control of High-Harmonic Generation by Tuning the Electronic Structure and Carrier Injection*. Nano Letters, **20**, 6215 (2020)

[27] C. H. Lui, K. F. Mak, J. Shan, and T. F. Heinz, *Ultrafast Photoluminescence from Graphene*. Phys. Rev. Lett. **105**, 127404 (2010).

[28] W.-T. Liu, S. W. Wu, P. J. Schuck, M. Salmeron, Y. R. Shen, and F. Wang, *Nonlinear broadband photoluminescence of graphene induced by femtosecond laser irradiation*. Phys. Rev. B **82**, 081408(R) (2010).

[29] S. Ono, *Thermalization in simple metals: Role of electron-phonon and phonon-phonon scattering*, Phys. Rev. B **97**, 054310 (2018).

[30] S. Ono and T. Suemoto, *Ultrafast photoluminescence in metals: Theory and its application to silver*. Phys. Rev. B **102**, 024308 (2020).

[31] L. P. Kadanoff and G. Baym, *Quantum Statistical Mechanics: Green's Function Methods in Equilibrium and Nonequilibrium Problems*, 1st ed. (CRC Press, Boca Raton, 1962).

[32] J. Reimann, S. Schlauderer, C. P. Schmid, F. Langer, S. Baierl, K. A. Kokh, O. E. Tereshchenko, A. Kimura, C. Lange, J. Güdde, U. Höfer, and R. Huber, *Subcycle observation of lightwave-driven Dirac currents in a topological surface band.* Nature **562**, 396 (2018).

[33] S. Ghimire, A. D. DiChiara, E. Sistrunk, P. Agostini, L. F. DiMauro, and D. A. Reis, *Observation of high-order harmonic generation in a bulk crystal.* Nat. Phys. **7**, 138 (2011).

[34] Y. S. You, D. A. Reis, and S. Ghimire, *Anisotropic high-harmonic generation in bulk crystals*. Nat. Phys. **13**, 345 (2017).

[35] H. Liu, Y. Li, Y. S. You, S. Ghimire, T. F. Heinz, and D. A. Reis, *High-harmonic generation from an atomically thin semiconductor*. Nat. Phys. **13**, 262 (2017).

[36] F. Langer, M. Hohenleutner, U. Huttner, S. W. Koch, M. Kira, and R. Huber, *Symmetry-controlled temporal structure of high-harmonic carrier fields from a bulk crystal*. Nat. Photonics **11**, 227 (2017).

[37] T. T. Luu and H. J. Wörner, *Measurement of the Berry curvature of solids using high-harmonic spectroscopy*. Nat. Commun. **9**, 916 (2018).

[38] H. Lakhotia, H. Y. Kim, M. Zhan, S. Hu, S. Meng, and E. Goulielmakis, *Laser picoscopy of valence electrons in solids*. Nature **583**, 55 (2020).

[39] C. P. Schmid, L. Weigl, P. Grössing, V. Junk, C. Gorini, S. Schlauderer, S. Ito, M. Meierhofer, N. Hofmann, D. Afanasiev, J. Crewse, K. A. Kokh, O. E. Tereshchenko, J. Güdde, F. Evers, J. Wilhelm, K. Richter, U. Höfer, and R. Huber, *Tunable non-integer high-harmonic generation in a topological insulator*. Nature **593**, 385 (2021).

[40] K. Uchida, V. Pareek, K. Nagai, K. M. Dani, and K. Tanaka, *Visualization of two-dimensional transition dipole moment texture in momentum space using high-harmonic generation spectroscopy*. Phys. Rev. B **103**, L161406 (2021).

# Supplementary information

K. S. Takeda[1], K. Uchida[1,*], K. Nagai[1], S. Kusaba[1], and K. Tanaka[1,*]

[1] *Department of Physics, Graduate School of Science, Kyoto University, Kyoto, Kyoto 606-8502, Japan*

*e-mail: uchida.kento.4z@kyoto-u.ac.jp, kochan@scphys.kyoto-u.ac.jp

## S1.  Sample preparation and characterization

We prepared thin layers 2H-NbSe$_2$ by mechanical exfoliation of a commercial 2H-NbSe$_2$ crystal grown by chemical vapor deposition (CVD), which was purchased from HQ-graphene Inc.. The thin layers were transferred onto the fused silica substrate by dry transfer method. Figure S1 shows optical microscope image of the sample on the substrate mainly used in HHG measurements. The sample thickness was characterized by reflection spectroscopy and atomic force microscopy. Crystal orientation of the sample was determined by second harmonic generation (SHG) spectroscopy [1].

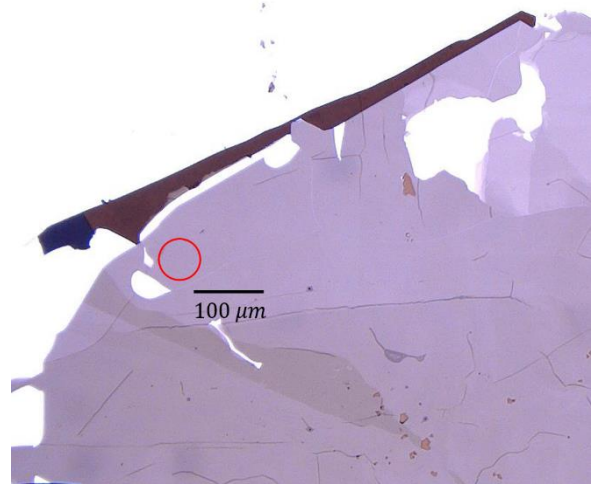

FIG. S1 Optical image of the exfoliated 2H-NbSe2 on substrate. Red circle indicates the MIR spot mainly used in HHG measurements.

## S2. Band structure of 2H-NbSe$_2$

For the simulation of HHG response and electron distribution, we calculated band structure of 2H-NbSe$_2$. The band structure of transition metal dichalcogenides near the Fermi energy level is mainly composed of three d-orbitals of chalcogen atom, and dispersion of these bands is quasi two-dimensional. Hence, we adopt the band structure calculated by using the two-dimensional tight binding model with three d-orbitals [2,3]. Figure S2 shows the calculated three bands, and we chose the band that crosses the Fermi energy as $E(\boldsymbol{k})$ in the main text because we ignore the interband tunneling process.

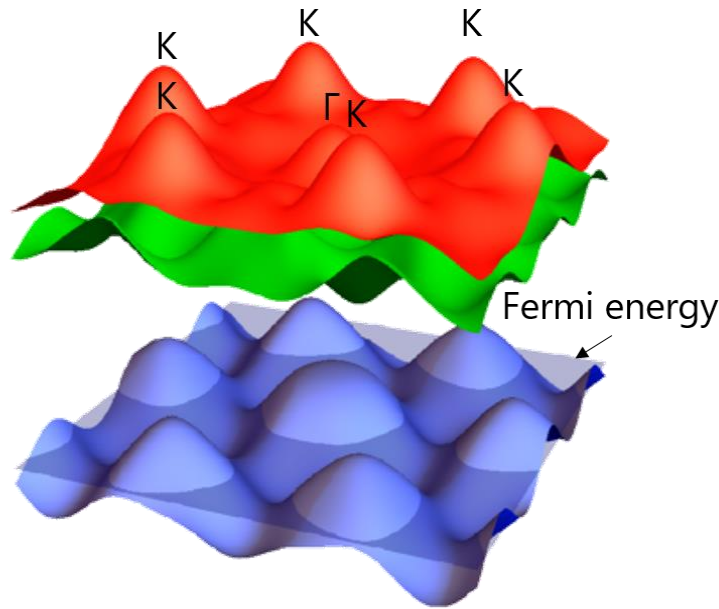

FIG. S2 Calculated band structure of 2H-NbSe2 using the model and parameters in Ref. 3.

## S3. Theoretical calculation

Here, we consider the electron dynamics based on the Boltzmann transport equation [4,5]:

$$\frac{\partial f}{\partial t} + \frac{1}{\hbar}\frac{\partial E(\boldsymbol{k})}{\partial \boldsymbol{k}} \cdot \frac{\partial f}{\partial \boldsymbol{x}} - \frac{e}{\hbar}\frac{\partial U_{\text{eff}}}{\partial \boldsymbol{x}} \cdot \frac{\partial f}{\partial \boldsymbol{k}} = \Sigma_{\text{scatter}}, \quad (4)$$

where $f = f(\boldsymbol{r}, \boldsymbol{k}, t)$ is the statistical distribution function of electron ensemble, $E(\boldsymbol{k})$ is the band structure, $U_{\text{eff}}(\boldsymbol{x}, t)$ is the effective potential, $e > 0$ is the elementary charge and $\hbar$ is the Dirac constant.

The dimensionless form of eq.(4) is given by

$$\frac{\partial f}{\partial \tau} + \frac{a\Delta_t\Delta_E}{\hbar\Delta_x}\frac{\partial \epsilon(\boldsymbol{\kappa})}{\partial \boldsymbol{\kappa}} \cdot \frac{\partial f}{\partial \boldsymbol{\chi}} - \frac{ea\Delta_t\Delta_U}{\hbar\Delta_x}\frac{\partial u_{\text{eff}}}{\partial \boldsymbol{\chi}} \cdot \frac{\partial f}{\partial \boldsymbol{\kappa}} = \Delta_t\Sigma_{\text{scatter}}. \quad (5)$$

Here, for dimensionless treatment, we took $a$ as the lattice constant, $\Delta_t$ as the pulse duration (60 fs), $\Delta_x$ as the beam spot size (60 μm), $\Delta_E$ as the band width (1 eV), and $\Delta_U$ as the amplitude of the potential. $\Delta_U$ is equal to the $\Delta_x\Delta_{\mathcal{E}}$, where $\Delta_{\mathcal{E}}$ is the amplitude of the electric field (~ 1 MV/cm). Calculating the dimensionless parameters in eq.(5), $a\Delta_t\Delta_E/\hbar\Delta_x$ is $\sim 10^{-3}$ and $ea\Delta_t\Delta_U/\hbar\Delta_x$ is $\sim 1$. Hence, we can ignore the second term of eq.(4), and the spatial dependence of the Boltzmann equation is eliminated as follows:

$$\frac{\partial f}{\partial t} + \frac{e}{\hbar}\mathcal{E}(t) \cdot \frac{\partial f}{\partial \boldsymbol{k}} = \Sigma_{\text{scatter}}. \quad (6)$$

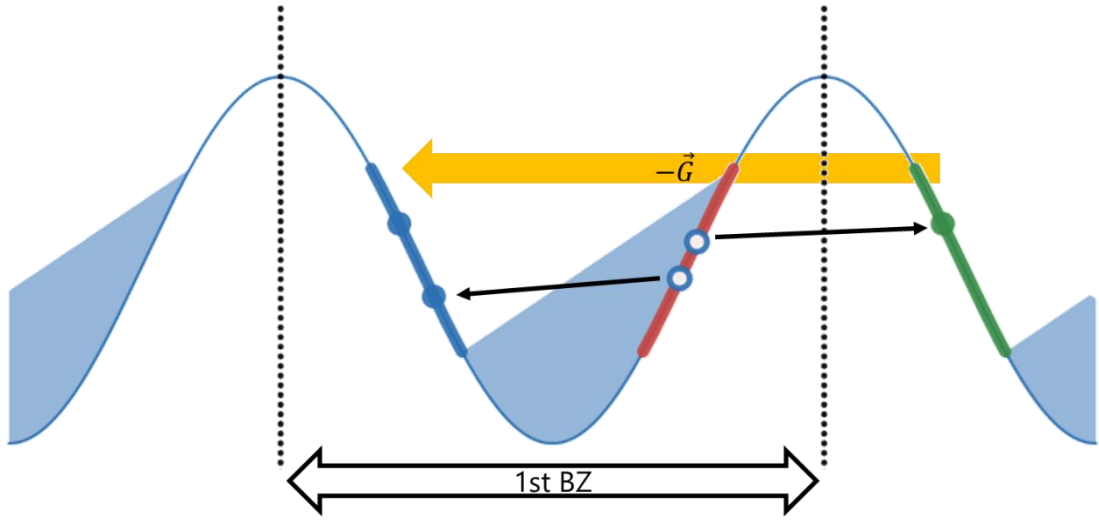

The electron-electron (e-e) scattering term is expressed by the following form [3]:

$$\int_{BZ^3} s_{ee}(\boldsymbol{k}, \boldsymbol{k}', \boldsymbol{k}_1, \boldsymbol{k}_1') \\ \times \{f'f_1'(1-f)(1-f_1) - ff_1(1-f')(1-f_1')\} dk' dk_1 dk_1', \tag{7}$$

where $s_{ee}(\boldsymbol{k}, \boldsymbol{k}', \boldsymbol{k}_1, \boldsymbol{k}_1')$ is the scattering cross section. This term makes the eq.(6) difficult to solve due to the nonlocal nonlinearity.

FIG. S3 Schematic diagram of e-e scattering in 1D cosine band. Blue shaded area represents the electron distribution in reciprocal space driven by the electric field. Red line shows the region where electrons can be scattered without being prohibited by Pauli blocking. Two blue open circles show the initial positions of electrons. Blue and Green bold line show the regions to which electrons in red region are scattered. Blue and Green solid circle indicated by arrows show the scattered electrons. Yellow arrow shows the reciprocal lattice vector. The region between the dotted lines shows the first Brillouin zone.

To approximate the scattering term into a simpler form, we firstly consider how e-e scattering occurs in the 1D cosine band. Figure S3 shows the schematic diagram of the e-e scattering process. When the electrons are driven by the electric field, electron distribution function $f(\boldsymbol{k},t)$ moves in reciprocal space. Because energy and momentum must be conserved in e-e scattering, electrons in the red region can be scattered to the blue or green regions in Fig. S3. Since the crystal momenta shifted by the reciprocal lattice vector are equivalent to each other, electrons scattered in the green region can be considered to be scattered in the blue region. This process is so-called the Umklapp process, and we can effectively regard the e-e scattering as generalized back-scattering form in 1D cosine band. Thus, we assumed the scattering form as follows [6]:

$$\Sigma_{\text{scatter}} = \sum_g \gamma_g \{f(\boldsymbol{k},t) - f(g\boldsymbol{k},t)\}, \tag{8}$$

where g is a symmetric operation in 2H-NbSe$_2$ and $\gamma_g$ is the scattering rate of each scattering paths. Note that there are other the scattering paths in more complicated and two-dimensional electronic band structure. Here, we assumed that the contributions of other scattering paths are minor for simplicity.

The Boltzmann transport equation with the above phenomenological e-e scattering term is still difficult to solve numerically because it is nonlocal with respect to wavenumber. To save the computational cost and keep the accuracy for HHG calculation, we performed the numerical calculations by the following method. First, we substituted the Fourier series expansion of the distribution $f(\mathbf{k}, t)$:

$$f(\mathbf{k}, t) = \sum_{\mathbf{R}} f_{\mathbf{R}}(t) e^{i\mathbf{k}\cdot\mathbf{R}}, \qquad (9)$$

into the Boltzmann equation:

$$\frac{\partial f}{\partial t} + \frac{e}{\hbar}\mathcal{E}(t) \cdot \frac{\partial f}{\partial \mathbf{k}} = \sum_{g} \gamma_g \{f(\mathbf{k}, t) - f(g\mathbf{k}, t)\}, \qquad (10)$$

where $\mathbf{R}$ is a lattice vector, $e$ is the elementary charge, $\hbar$ is the Plank constant, $\mathcal{E}(t)$ is the MIR electric field, g is a symmetry operation in 2H-NbSe$_2$, $\gamma_g$ is the corresponding scattering rate. If $\gamma_{g_1}$ and $\gamma_{g_2}$ belong to the same conjugacy class of the crystallographic point group, then $\gamma_{g_1} = \gamma_{g_2}$. After substitution and some calculation, we can obtain the following simultaneous ordinary differential equations (SODE) labeled by $\mathbf{R}$:

$$\frac{df_{\mathbf{R}}(t)}{dt} + i\frac{e}{\hbar}\mathcal{E}(t) \cdot \mathbf{R} f_{\mathbf{R}}(t) = \sum_{g} \gamma_g \{f_{\mathbf{R}}(t) - f_{g\mathbf{R}}(t)\}. \qquad (11)$$

This SODE has a closed form for a set of lattice vectors that are transformed with symmetric operations to each other. Since the crystallographic point group of 2H-NbSe$_2$ is D$_{6h}$, there are 24 symmetric operations. Here, for simplicity, we only considered 12 in-plane symmetry operations by regarding the electronic structure as two-dimensional one. Since the one of the operation is the identity operation, there are 11 scattering channels at maximum as shown in Fig. S4.

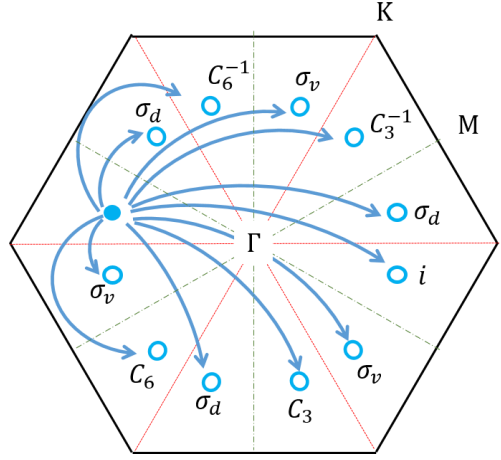

FIG. S4 Scattering channels of electron in two-dimensional Brillouin zone of 2H-NbSe$_2$. Filled circle indicates the electron before scattering. Open circles indicate the position of electron after scattering. The labels in open circles represent the corresponding symmetric operation in Eq. (5).

The initial condition is given by the Fourier expansion factor of the Fermi distribution. After calculating the time evolution of the Fourier coefficients of the $f(\bm{k},t)$ by (11), the nonlinear current can be calculated by

$$\bm{j}(t) = -e \int_{BZ} \frac{\partial E(\bm{k})}{\partial \bm{k}} f(\bm{k},t) d\bm{k}, \qquad (12)$$

where $E(\bm{k})$ is the electronic band structure. Substituting the Fourier series expansion of the band structure given by

$$E(\bm{k}) = \sum_{\bm{R}} E_{\bm{R}} e^{i\bm{k}\cdot\bm{R}}, \qquad (13)$$

and the Fourier series expansion of the distribution into (12), it becomes

$$j(t) = -e \sum_{R} R E_R \, \text{Im} \, f_{-R}(t). \tag{14}$$

If there is no scattering, eq.(11) can be solved easily, and the expression of current is given by following form:

$$j(t) = -e \sum_{R} R E_R f_{-R}(0) \exp\left(-i\frac{e}{\hbar} R \cdot \int_{-\infty}^{t} \mathcal{E}(t) dt\right) \tag{15}$$

In the simulation with e-e scattering, we set the cutoff for a lattice vector as $|R| < 50a$ by checking the convergence of the calculation for the ninth order harmonics. This numerical simulation of a finite set of SODE greatly saves the computational cost and also keeps the accuracy of simulated HHG results.

## S4. Estimation of the scattering rate

The total energy of electronic system is given by

$$E_{\text{total}}(t) = \int_{\text{BZ}} E(\boldsymbol{k}) f(\boldsymbol{k}, t) d\boldsymbol{k}. \tag{16}$$

After driving field has passed, the electron distribution is relaxed into the stationary distribution $f(\boldsymbol{k}, \infty)$ through e-e scattering. The corresponding total energy becomes

$$E_{\text{total}}(\infty) = \int_{\text{BZ}} E(\boldsymbol{k}) f(\boldsymbol{k}, \infty) d\boldsymbol{k}. \tag{17}$$

Note that $f(\boldsymbol{k}, \infty)$ and equilibrium distribution (the Fermi distribution) $f_{\text{F}}(\boldsymbol{k}, T)$ are not necessarily equal in our calculation, and we cannot define the temperature of electrons after excitation in the simulation due to the simplification of complicated e-e scattering process into back-scattering form discussed above. However, in real systems, if the energy flow out of the electronic system is negligible, the electron distribution should approach the equilibrium distribution whose total energy is the same as $f(\boldsymbol{k}, \infty)$. Hence, we determined the effective temperature of $f(\boldsymbol{k}, \infty)$ by using the following condition:

$$\int_{\text{BZ}} E(\boldsymbol{k}) f(\boldsymbol{k}, \infty) d\boldsymbol{k} = \int_{\text{BZ}} E(\boldsymbol{k}) f_{\text{F}}(\boldsymbol{k}, T) d\boldsymbol{k}. \tag{18}$$

Since $f(\boldsymbol{k},\infty)$ depends on scattering rate, we can estimate the scattering rate that relaxes to the thermal equilibrium distribution at a given temperature (See FIG. 2(c) in the main paper).

## S5. The effect of time-evolution of electron temperature on broadband emission spectrum

Here, we consider the effect of time-varying electron temperature on broadband emission spectrum. Through the coupling with the lattice degree of freedom, effective temperature of electrons gradually decreases in time, for example, as shown in Fig. S5. Therefore, the observed spectrum is the averaged one over the wide range of temperature from several thousand K to room temperature. To consider this effect, we assume that time evolution of effective electron temperature is given by the following (Fig. S5(a)):

$$T_{\text{eff}}(t) = T_0 + \Delta T \exp\left(-\frac{t}{\tau_p}\right). \quad (15)$$

Here, we set $T_0 = 300$ K, $\Delta T = 3700$ K, which is consistent with our experiment, and $\tau_p$ is the decay time of effective temperature. The thermal emission intensity $I_b(t)$ at time $t$ is given by

$$I_b(t) = I_b(T_{\text{eff}}(t), \hbar\omega) \propto \frac{\omega^3}{\exp\left(\frac{\hbar\omega}{k_B T_{\text{eff}}(t)}\right) - 1}. \quad (16)$$

The observed broadband emission intensity is the integrated one over time as follows:

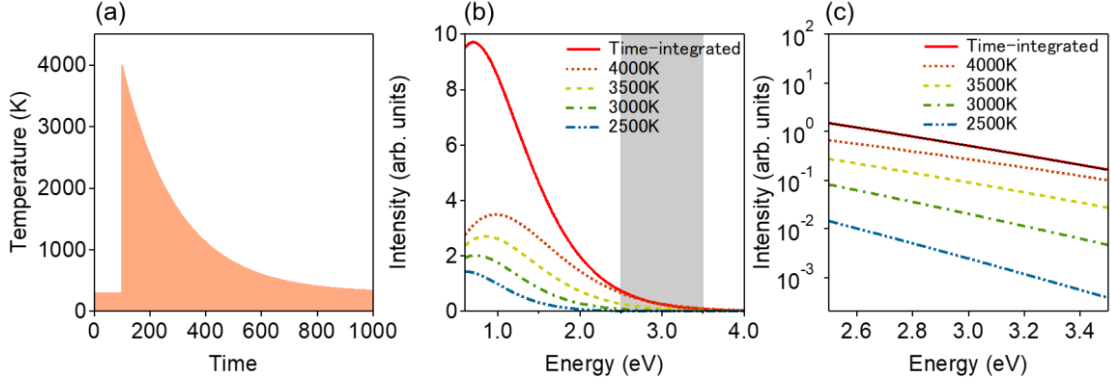

FIG. S5 (a) Temporal profile of electron temperature that we assumed in the simulation. The horizontal axis is arbitrary unit. (b) Time-integrated thermal emission spectrum with electron temperature dynamics shown in (a) (red solid line). Brown dotted, yellow dashed, green dashed-dotted, and blue dashed-double-dotted lines show thermal emission spectra with electron temperature of 4000 K, 3500 K, 3000 K, and 2500 K, respectively. The intensities of stationary thermal emission are corrected according to Eq. (17). Gray shaded area corresponds to the region where we performed the fitting in the main text. (c) The expanded traces of (b). Black solid line indicates the fitting result using Planck distribution function.

$$I_b(\hbar\omega) \propto \int ds\, I_b(T_{\text{eff}}(s), \hbar\omega) = \tau_p \int_{T_0}^{T_0+\Delta T} \frac{dT\, I_b(T, \hbar\omega)}{T - T_0}. \qquad (17)$$

The result indicates that the observed spectrum can be written by the integration over temperature with the weight of the inverse of temperature, and spectral shape does not depends on the decay time $\tau_p$. Figure S5(b) show the time-integrated thermal emission spectrum at the peak temperature of 4000 K (Red solid line) and steady-state thermal emission spectra at several temperatures. In the temperature range between room temperature and 4000 K, the peaks of spectrum of Planck distribution for each temperature are located in infrared region (below 1.5 eV). Therefore, the effect of time-varying temperature on thermal emission spectrum is only salient in the infrared region. In visible region, the contribution from the peak temperature (4000 K) is dominant as shown in Fig. S5(c). In fact, the time-integrated spectrum in visible range is well fitted by the Planck distribution with temperature of 3700 K. Based on this result, we conclude that the fitting of the broadband emission spectrum in visible region with Planck distribution yields the peak temperature value reported in main paper.

# S6. Propagation effect inside the sample

The thickness of the sample used in the experiment was about 25 nm, which seems to be thin enough for light propagation. However, since 2H-NbSe$_2$ is a metal, it is necessary to estimate the attenuation of the electromagnetic field, impedance mismatch, and phase matching. To estimate these quantities, we calculate the dielectric function obtained from the Kramers-Kronig transformation of reflectivity spectra [7-12]. Figure S6 shows the calculated absorption coefficient. In the range of photon energy shown in the graph, the absorption length is longer than $10^{-7}$ m, which is much longer than the thickness of the sample. Hence, we can neglect the depletion of the MIR intensity and reabsorption of the harmonics.

Figure S7 shows the transmission coefficient at the SiO$_2$-NbSe$_2$ boundary and Vacuum-NbSe$_2$ boundary. In the range of photon energy of harmonics from 5th to 9th order, transmission coefficient at each boundary don't change much. Hence, spectral shape of calculated harmonics is not modified largely by the boundary reflection. Figure S8 shows the calculated coherence length. In the range of photon energy shown in the graph, the coherence length is longer than $10^{-6}$ m, which is much longer than the thickness of the sample. Hence, we can ignore the effects of phase mismatch.

From the above results, the effect of propagation on high harmonics can be neglected, and the experimental and numerical results can be directly compared in our experimental condition.

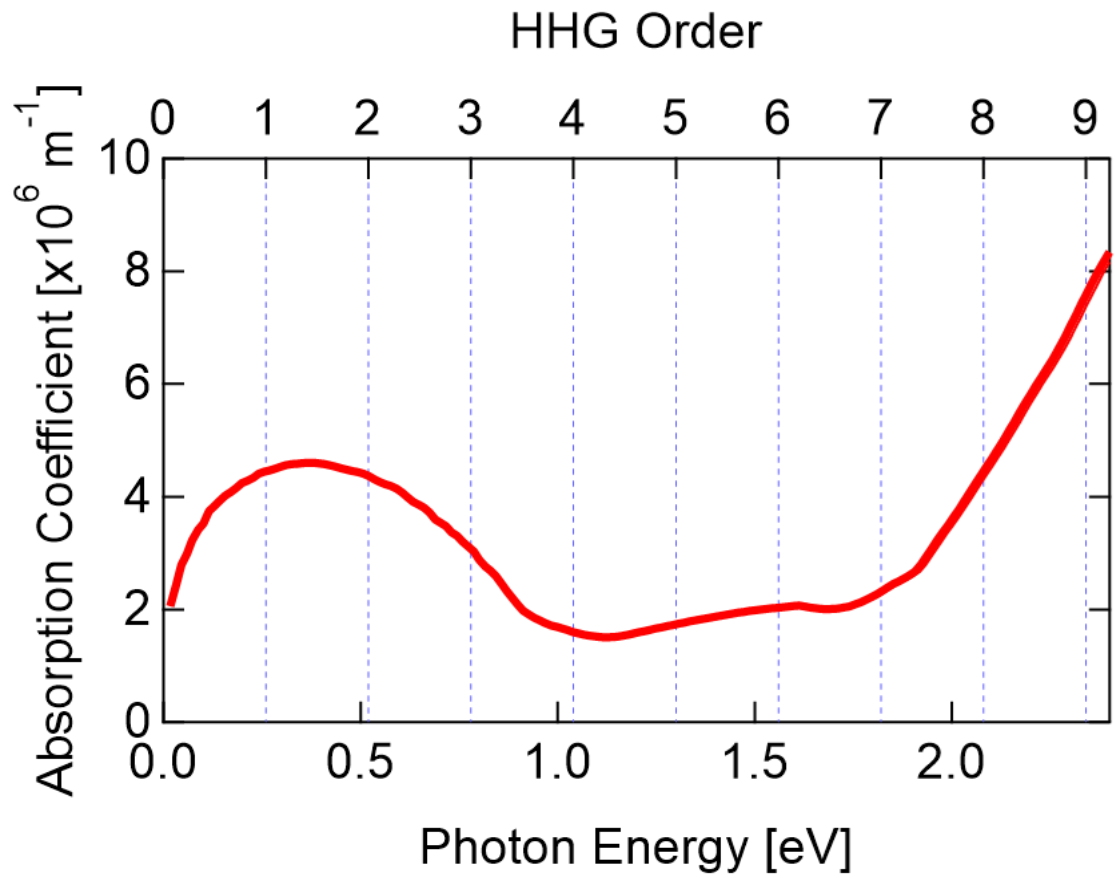

FIG. S6: Absorption coefficient of 2H-NbSe$_2$ in the range from 0 eV to 2.4 eV. The minimum value is 1.5 μm$^{-1}$ at 1.1 eV.

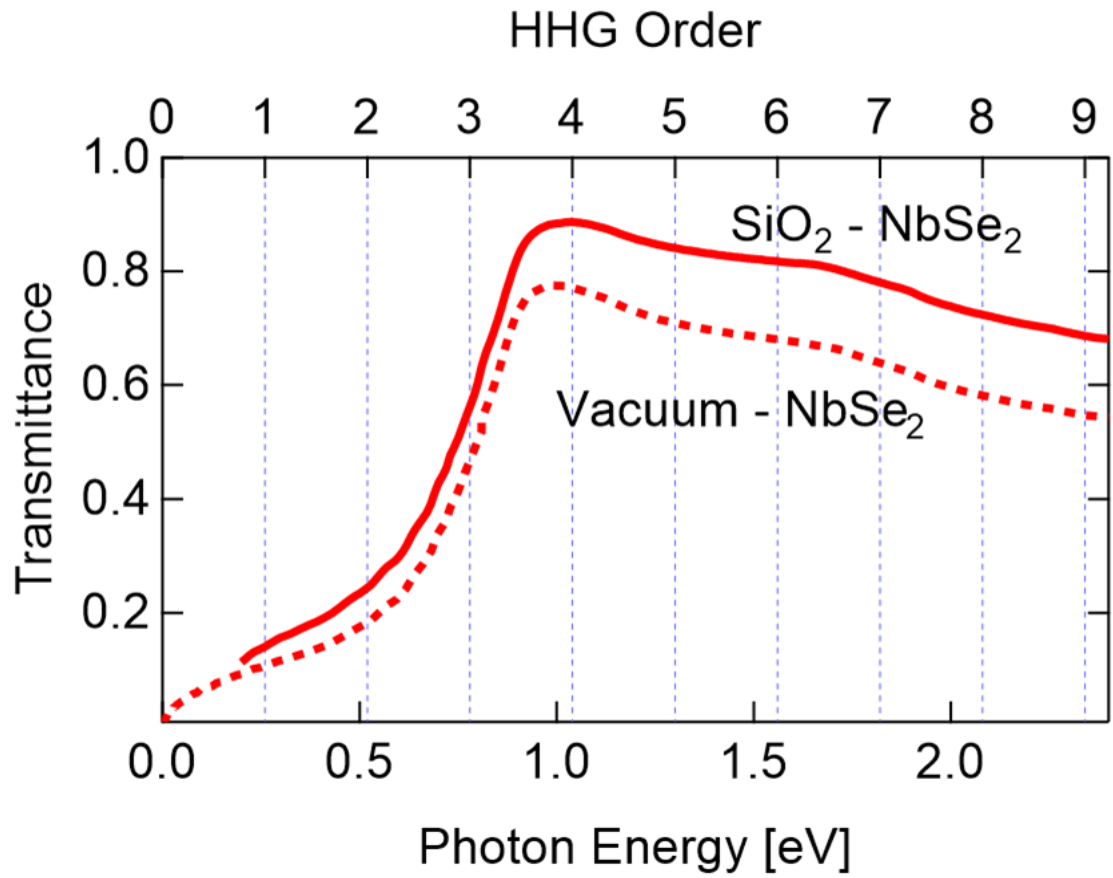

FIG. S7 Transmittance of 2H-NbSe$_2$ in the range from 0 eV to 2.4 eV. Solid line represents the that at SiO$_2$-NbSe$_2$ boundary and break line shows the that at Vacuum-NbSe$_2$ boundary.

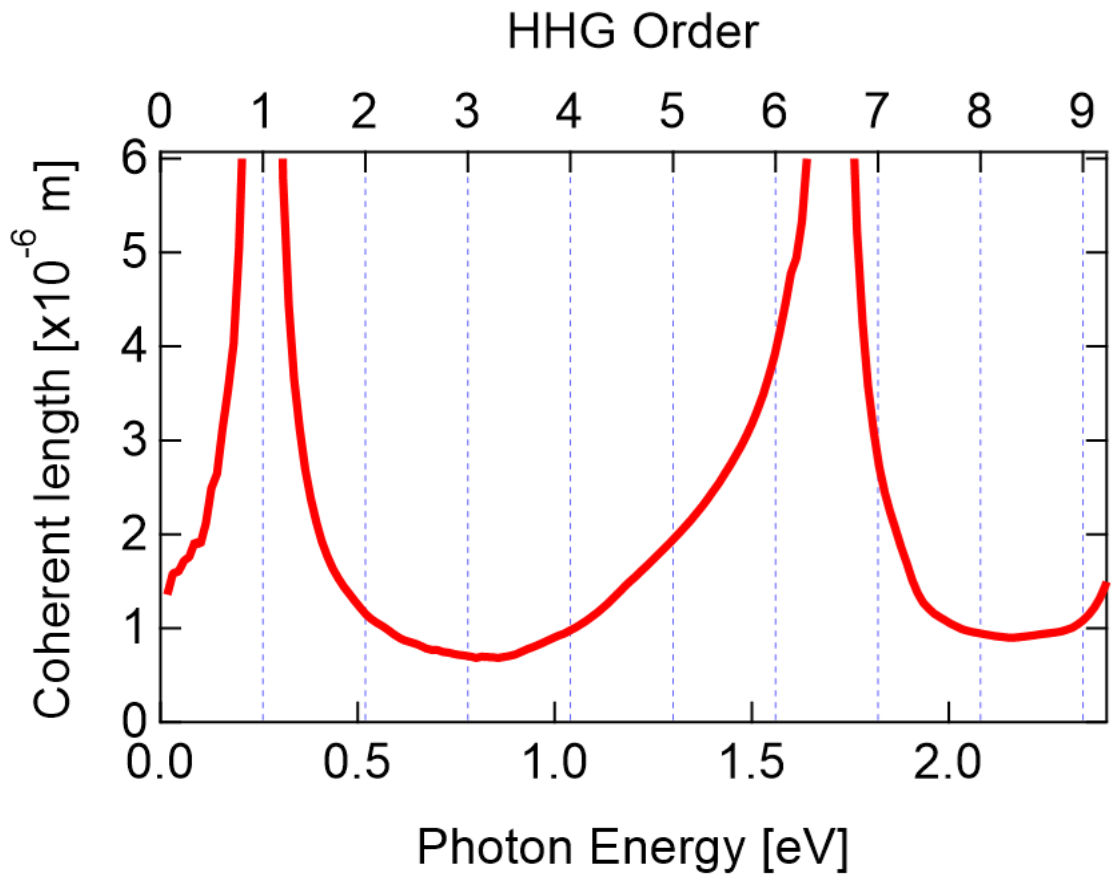

FIG. S8 Coherence length of light in the range from 0 eV to 2.4 eV in 2H-NbSe$_2$. The minimum value is about 685 nm at photon energy of 0.8 eV.

# S7.  Effect of the electron distribution on HHG anisotropy

Based on the intraband current model, nonlinear current which emit the high harmonics is the integrated one that takes contributions from all electrons below the Fermi level. Therefore, the distribution of electrons in reciprocal lattice space and the Fermi energy may affect the properties of HHG. To investigate this effect, we perform numerical calculations for different electron distributions and Fermi energies.

Figures S9 shows the numerical results of the crystal orientation dependence of seventh harmonics in different electron distributions. As in Fig. S9 (a), HHG contributions near $\Gamma$ point (right-hand side panel) are more anisotropic than those near K point (center panel), and integrated HHG response is more likely to those near K point (left-hand side panel). This clearly indicates that the crystal orientation dependence of high harmonics reflects the details of electron distribution. Figure S9(b) show the crystal orientation dependence of the seventh harmonics with several Fermi levels. The direction of MIR electric field where HHG yields has maximum value is shifted by changing Fermi level.  In semiconductors, anisotropic HHG response is considered to originate from anisotropic electronic structures. On the other hand, in metals, we found that we also need to consider the contribution of electron distribution. This may lead to precise control of extreme nonlinear optical properties by tuning Fermi level and electron distribution.

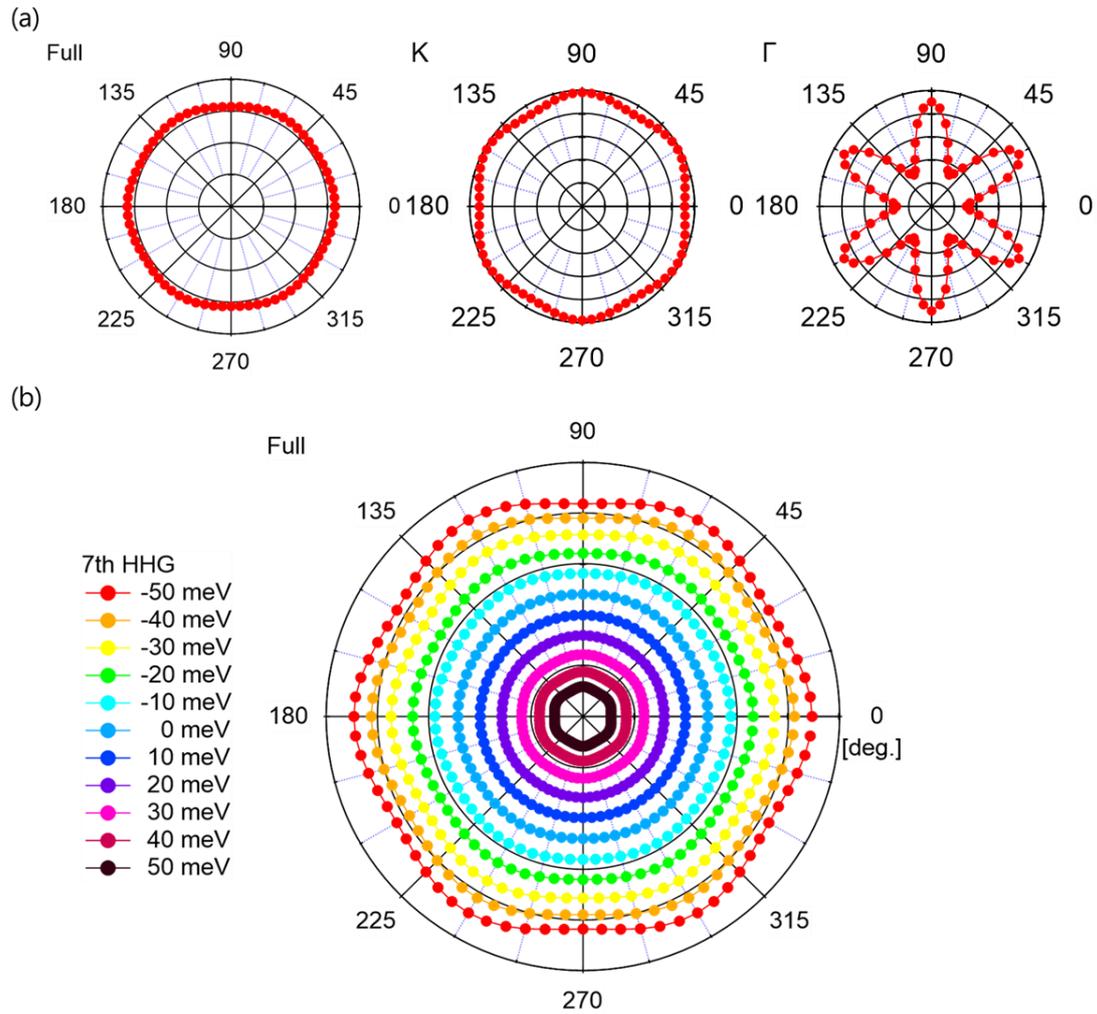

FIG. S9 The change of the crystal orientation dependence of 7th harmonics with respect to the electron distribution and the Fermi energy. The left of (a) shows the case when all electrons are considered, the middle shows the case when only electrons in K and K' point are considered, and the right shows the case when only electron in Γ point are considered. (b) shows the change of the crystal orientation dependence of 7th harmonics when the fermi energy is changed.

## S6. Power dependence of the broadband emission

Figure S6 shows the MIR intensity dependence of broadband emission intensity integrated from 1.5 eV to 4.0 eV. It depends nonlinearly on the excitation light intensity, proportional to the fifth power of the excitation light intensity when the excitation light intensity is weak, and proportional to the third power when the excitation light intensity is strong. This result suggests that non-perturbative nonlinearity is involved in broadband emission process.

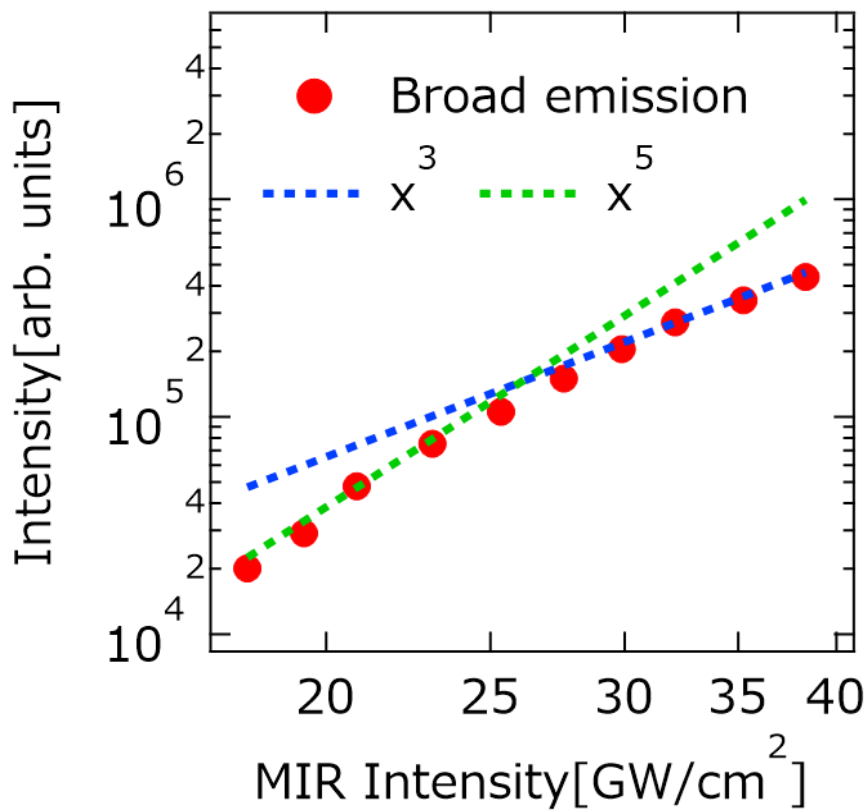

FIG. S6 The MIR intensity dependence of broad emission. The intensity of broad emission is integrated from 1.5 eV to 4.0 eV. Red circles show the integrated intensity of broad emission. Blue break line and green break line show extrapolation lines for proportional to the 3th and 5th power of intensity, respectively.

[1] X. Xi, Z. Wang, W. Zhao, J.-H. Park, K. T. Law, H. Berger, L. Forró, J. Shan, and K. F. Mak, *Ising pairing insSuperconducting NbSe$_2$ atomic layers*. Nat. Phys. **12**, 139 (2016).

[2] G.-B. Liu, W.-Y. Shan, Y. Yao, W. Yao, and D. Xiao, *Three-band tight-binding model for monolayers of group-VIB transition metal dichalcogenides*. Phys. Rev. B **88**, 085433 (2013).

[3] D. Möckli and M. Khodas, *Robust parity-mixed superconductivity in disordered monolayer transition metal dichalcogenides*. Phys. Rev. B **98**, 144518 (2018).

[4] L. P. Kadanoff and G. Baym, *Quantum Statistical Mechanics: Green's Function Methods in Equilibrium and Nonequilibrium Problems*, 1st ed. (CRC Press, Boca Raton, 1962).

[5] A. Jüngel, *Transport Equations for Semiconductors*. 1st ed. (Springer, Berlin, Heidelberg, 2009).

[6] J. Reimann, S. Schlauderer, C. P. Schmid, F. Langer, S. Baierl, K. A. Kokh, O. E. Tereshchenko, A. Kimura, C. Lange, J. Güdde, U. Höfer, and R. Huber, *Subcycle observation of lightwave-driven Dirac currents in a topological surface band.*, Nature **562**, 396 (2018).

[7] R. Bachmann, H. C. Kirsch, and T. H. Geballe, *Optical properties and superconductivity of NbSe$_2$*. Solid State Commun. **9**, 57 (1971).

[8] W. Y. Liang, *Reflectivity of MoS$_2$ and NbSe$_2$ (Interband transitions)*. J. Phys. C: Solid State Phys. **4**, L378 (1971).

[9] G. Leveque, S. Robin-Kandare, L. Martin, and F. Pradal, *Reflectivity of MoS$_2$ and NbSe$_2$ in the Extreme UV Range (20 to 70 eV)*. Phys. Status Solidi B **58**, K65 (1973).

[10] L. Martin, R. Mamy, A. Couget, and C. Raisin, *Optical Properties and Collective Excitations in MoS$_2$ and NbSe$_2$ in the 1.7 to 30 eV Range.* Phys. Status Solidi B **58**, 623 (1973).

[11] A. R. Beal, H. P. Hughes, and W. Y. Liang, *The reflectivity spectra of some group VA transition metal dichalcogenides.* J. Phys. C: Solid State Phys. **8**, 4236 (1975).

[12] S. V. Dordevic, D. N. Basov, R. C. Dynes, and E. Bucher, *Anisotropic electrodynamics of layered metal 2H-NbSe$_2$.* Phys. Rev. B **64**, 161103(R) (2001).